\documentclass[aps,prd,nofootinbib,amsmath,amssymb,showpacs,longbibliography,singlecolumn,superscriptaddress]{revtex4-2}

\usepackage{epsfig}
\usepackage{bm}
\usepackage{amssymb}
\usepackage{amsmath}
\usepackage{color}
\usepackage{subfigure}
\usepackage{dcolumn}

\usepackage[utf8]{inputenc}

\usepackage{amsmath}
\usepackage{amssymb}
\usepackage{amsthm}
\usepackage{amscd, amsfonts, mathrsfs}
\usepackage{cases}
\usepackage{verbatim} 
\usepackage{epsf}
\usepackage[bookmarksnumbered,pdfpagelabels=true,plainpages=false,colorlinks=true,
            linkcolor=black,citecolor=black,urlcolor=black]{hyperref}

\theoremstyle{plain}

\theoremstyle{definition}

\allowdisplaybreaks

\newcommand{\be}{\begin{equation}}
\newcommand{\ee}{\end{equation}}
\newcommand{\bea}{\begin{eqnarray}}
\newcommand{\eea}{\end{eqnarray}}

\newcommand{\bml}{\begin{subequations}}
\newcommand{\eml}{\end{subequations}}

\newcommand{\bbm}{\begin{bmatrix}}
\newcommand{\ebm}{\end{bmatrix}}
\newcommand{\bvm}{\begin{vmatrix}}
\newcommand{\evm}{\end{vmatrix}}

\begin{document}


\title{Far-from-equilibrium kinetic dynamics of $\lambda \phi^4$ theory in an expanding universe}
\date{\today}

\author{Nicki Mullins}
\email{nickim2@illinois.edu}
\affiliation{Illinois Center for Advanced Studies of the Universe\\ Department of Physics, 
University of Illinois at Urbana-Champaign, Urbana, IL 61801, USA}


\author{Gabriel Denicol}
\email{gsdenicol@id.uff.br}
\affiliation{Instituto de F\'isica, Universidade Federal Fluminense,
Av.\ Milton Tavares de Souza, Niter\'oi, Brazil, Zip Code: 24210-346}

\author{Jorge Noronha}
\email{jn0508@illinois.edu}
\affiliation{Illinois Center for Advanced Studies of the Universe\\ Department of Physics, 
University of Illinois at Urbana-Champaign, Urbana, IL 61801, USA}


\begin{abstract}
We investigate the far-from-equilibrium behavior of the Boltzmann equation for a gas of massless scalar field particles with quartic (tree level) self-interactions ($\lambda \phi^4$) in Friedmann-Lemaitre-Robertson-Walker spacetime. Using a new covariant generating function for the moments of the Boltzmann distribution function, we analytically determine a subset of the spectrum and the corresponding eigenfunctions of the linearized Boltzmann collision operator. We show how the covariant generating function can be also used to find the exact equations for the moments in the full nonlinear regime. Different than the case of a ultrarelativistic gas of hard spheres (where the total cross section is constant), for $\lambda \phi^4$ the fact that the cross section decreases with energy implies that moments of arbitrarily high order directly couple to low order moments. Numerical solutions for the scalar field case are presented and compared to those found for a gas of hard spheres.     
\end{abstract}

\maketitle


\section{Introduction}

The Boltzmann equation plays an important role in our understanding of the properties of dilute gases \cite{Cercignani}. In the relativistic regime, the Boltzmann equation \cite{degroot} has been widely applied to describe phenomena in many different fields, ranging from cosmology \cite{Bernstein:1988bw} to heavy-ion collisions \cite{Heinz:1984yq,Heinz:1985qe,Bass:1998ca,Arnold:2000dr,Xu:2004mz,Denicol:2012cn,Denicol:2019lio}. This nonlinear integro-differential equation describes how the single-particle distribution function evolves in phase-space in the presence of collisions among the constituents of the gas. From the single-particle distribution function quantities associated with conservation laws such as the energy-momentum tensor, or the particle current, can be reconstructed, which provides a way to determine the hydrodynamic evolution of the system.  

The properties of the non-relativistic Boltzmann equation have been widely investigated \cite{Cercignani}. In fact, besides extensive numerical solutions, an exact solution for a non-relativistic gas of Maxwell molecules has been derived \cite{Bobylev,KrookWu1976,KrookWu1977}. This result, known as the Bobylev-Krook-Wu solution, was found by deriving an exact set of coupled nonlinear differential equations for the moments of the distribution function, which then admit an analytical solution for a given choice of initial conditions. Following along the work by Bobylev, Krook, and Wu, in \cite{Bazow:2015dha, Bazow:2016oky} the relativistic Boltzmann equation for a gas of particles interacting with constant cross section, in an expanding spacetime, was rewritten in terms of an exact (infinite) set of coupled differential equations for moments of the single-particle distribution function. In that case, because the cross section is independent of the energy, the equations for the moments can be solved recursively, with the solution of moments of order $n$ only depending on moments of order $k < n$. This property was crucial to find one analytical solution for the moments, which in turn led to the first (and so far, only) analytical solution of the full nonlinear 
Boltzmann equation for a relativistically expanding gas \cite{Bazow:2015dha}. This exact solution was then compared to approximate solutions of the Boltzmann equation obtained by employing the relaxation time approximation \cite{ANDERSON1974466,Rocha:2021zcw} and the linearized collision term, which was useful to determine the validity of such approximations \cite{Bazow:2016oky}.

In this paper we go beyond \cite{Bazow:2015dha, Bazow:2016oky} and consider the Boltzmann equation for a gas of classical massless scalar field particles with quartic (tree-level $\lambda \phi^4$) self-interactions  in a homogeneously expanding, isotropic spacetime. A new covariant generating function method for the moments of the Boltzmann distribution function is introduced in this work to analytically determine the  subset of scalar eigenfunctions of the linearized Boltzmann collision operator, and the corresponding  eigenvalues. This covariant generating function is then employed to find the exact equations of motions for the scalar moments in the full nonlinear regime. Unlike the case of hard spheres, for $\lambda \phi^4$ the cross section decreases with energy  and, as shall be demonstrated in this paper, this implies that moments of all orders are coupled to each order. This coupling prevents finding a full analytical solution, but numerical solutions of these moment equations can be obtained using simple numerical schemes. A comparison between such numerical solutions, and the corresponding solutions found for a gas of hard spheres with the same initial conditions, is presented in this work.

This paper is organized as follows. In Section \ref{boltzmann} we discuss the general properties of the Boltzmann equation in Friedmann-Lemaitre-Robertson-Walker spacetime. In Section \ref{linearized_soln} we introduce the generating function method and use it to analytically derive the scalar subset of the spectrum and eigenfunctions of the linearized Boltzmann operator for $\lambda \phi^4$ theory. The scalar moment equations of the nonlinear Boltzmann equation for a scalar field are derived in Section \ref{moments_method}. Section \ref{numerics} includes some numerical results and comparisons between the scalar field and constant cross section solutions. A summary of our findings and the conclusions we have drawn are presented in section \ref{conclusion}. Technical details concerning the properties of associated Laguerre polynomials are given in Appendix \ref{Laguerre_properties}. In Appendix \ref{const_cross_soln} we show how the new generating function method introduced in this paper can be used to derive the exact moment equations for the constant cross section case, which were originally obtained in \cite{Bazow:2015dha,Bazow:2016oky} using different methods.

\emph{Notation}: We use a mostly minus metric and natural units. Four-vectors are defined as $a^\mu = (a^0,\mathbf{a})$ and we use $\cdot$ for the scalar product between  spatial vectors, i.e., $a_i b^i=\mathbf{a} \cdot \mathbf{b}$. 

\section{Boltzmann Equation in expanding spacetime}
\label{boltzmann}

We consider a homogeneous and isotropically expanding system of massless scalar particles with quartic self-interactions, embedded in a curved spacetime described by the Friedmann-Lemaitre-Robertson-Walker (FLRW) metric \cite{Weinberg_GR_book} (with zero spatial curvature). In the \emph{conformal gauge}  the line element is 
\begin{equation}
ds^{2}=a^{2}(\tau)\left(d\tau^{2}-dx^{2}-dy^{2}-dz^{2}\right) \,.  
\label{metric}
\end{equation}%
We note that the FLRW metric $g_{\mu\nu}$ written above is related to the standard Minkowski metric via Weyl rescaling \cite{Weinberg_GR_book}. This fact will play an important role when solving the Boltzmann equation in this curved spacetime, as we explain below. Note that in these coordinates the fluid 4-velocity is $u^{\mu }=\left(1/a(\tau),0,0,0\right) $, and the expanding FLRW geometry induces a nonzero fluid expansion rate $\theta (\tau)=\nabla_\mu u^\mu= \partial _{\mu }(\sqrt{-g}\,u^{\mu })/\sqrt{-g}=3Da/a^2$, where $u^\mu \nabla_\mu = D$ and  $\sqrt{-g}=a^{4}(\tau)$, with $g$ being the determinant of the metric in \eqref{metric}. Furthermore, for the FLRW metric above there are many nonzero Christoffel symbols,  all of them equal to $Da(\tau)/a(\tau)$. We take the probe limit in which the energy and momentum of the kinetic particles are negligible in comparison to other sources that define the underlying cosmological scale factor $a(\tau)$ of the metric. In this limit our results are valid for any $a(\tau)>0$ (for instance, for a radiation dominated universe, $a(\tau) \sim \tau$). 

The dynamics of the single-particle distribution function, $f(x,k)$, is given by the relativistic Boltzmann equation in curved space \cite{2009PhyA..388.1079D, 2009PhyA..388.1818D,Denicol:2014xca,Denicol:2014tha,Tinti:2018qfb}
\begin{equation}
k^{\mu }\partial _{\mu }f(x,k)+\Gamma _{\mu i}^{\lambda }k_{\lambda} k^{\mu } 
\frac{\partial f(x,k)}{\partial k_{i}}=\mathcal{C}[f].
\label{Boltzmann1}
\end{equation}
For massless particles with momentum $k^\mu$, the on-shell condition  $k_\mu k^\mu=0$ implies that $k^0 = |\mathbf{k}|=\sqrt{k_x^2 + k_y^2+k_z^2}$ (we only use covariant momenta). Since the FLRW universe is spatially homogeneous and isotropic, the distribution function must be homogeneous in space and only depend on the momentum via $u_\mu k^\mu$. Thus, we write $f(x,k)=f_k(\tau)$ from here on.

The symmetries of the FLRW spacetime strongly constrain the form of the conserved currents of the matter. Due to local momentum isotropy, the viscous shear-stress tensor, energy diffusion, and particle diffusion current vanish exactly. Therefore, for the massless gas one may write the energy-momentum tensor, $T^{\mu \nu }$, as 
\be
T_{\mu\nu} = \varepsilon\left( u_\mu u_\nu - \frac{1}{3}\Delta_{\mu\nu}\right)
\ee
and the particle 4-current, $N^{\mu }$, as
\be
N^\mu = n u^\mu.
\ee
Above, we introduced the spatial projector orthogonal to the 4-velocity, $\Delta_{\mu\nu} \equiv g_{\mu\nu}-u_\mu u_\nu$. In FLRW, the total energy  $\varepsilon = T^{\mu\nu}u_\mu u_\nu$ and particle $n = u_\mu N^\mu$ densities in the local rest frame are only functions of $\tau$. The time evolution of these quantities is fully determined by the conservation laws
\begin{equation}
Dn+3n\frac{Da}{a}=0,\qquad D\varepsilon +4\varepsilon \frac{Da}{a}=0.
\label{eqsenergydensity}
\end{equation}
With initial condition $a(\tau_{0}) = 1$, they are solved by $n(\tau) = {n_0}/a^{3}(\tau)$ and $\varepsilon (\tau) = \varepsilon_0/a^{4}(\tau)$, where $n_0$ and $\varepsilon_0$ are constants. Furthermore, for a conformal gas, one may write $\varepsilon \sim T^4$ and $n\sim T^3$ in terms of a suitably defined temperature $T$, such that $T(\tau) = T_0/a(\tau)$ and $T_0$ is the initial temperature scale.

As discussed in detail in \cite{Denicol:2014tha}, for a conformal system the Boltzmann equation transforms covariantly under a Weyl transformation of the metric $g_{\mu\nu} \to e^{-2\Omega}g_{\mu\nu}$. This is the case for the massless (on-shell) $\lambda \phi^4$ theory considered in this paper, where the corresponding interaction cross section does not break conformal invariance. This property implies that, if one conveniently writes the metric in conformal gauge, the factors of $a(\tau)$ will cancel and one can just solve the Boltzmann equation on the flat piece of \eqref{metric}. We will use this property to perform all of our calculations in Minkowski spacetime, leading to a fluid 4-velocity that is static, $u^\mu = (1,0,0,0)$. To recover the time dependence effect from the metric one can just use the well-known rules based on Weyl rescaling to obtain the quantities at hand (e.g., if $\varepsilon_0$ is the energy density computed using the flat dynamics, the energy density after recovering the Weyl factor will be simply $\varepsilon_0/a(\tau)^4$) \cite{Denicol:2014tha,Denicol:2014xca}. Another useful consequence is that the left-hand side of \eqref{Boltzmann1} considerably simplifies and the Boltzmann equation becomes 
\be
E_k\,D f_k(\tau) = \mathcal{C}[f],
\label{Boltzmanneq_general}
\ee
 where we defined the scalar $E_k = u_\mu k^\mu$. Also, after the Weyl rescaling of the metric, we rescale all the momenta ($k^\mu \to k^\mu/T_0$) and time ($\tau \to \tau T_0$) so both sides of the Boltzmann equation written above are dimensionless. We will work with those rescaled (dimensionless) quantities throughout the paper, unless otherwise specified.

In this work we will only consider the classical limit (Boltzmann statistics) where the Boltzmann equation for on-shell massless scalar particles with a $\lambda \phi^4$ interaction (i.e., a cross section $\sigma(s)\sim \lambda/s$, where $s$ is the square of the center of mass energy) can be written as \cite{degroot}
\bea
&&E_k\,D f_k =\mathcal{C}[f]= \frac{\lambda}{2}\,\int_{k'pp'}
\,(2\pi)^5\,\,\delta^{(4)}(k_\mu+k'_\mu-p_\mu -p'_\mu)(f_p f_{p'}{-}f_k f_{k'}),
\label{Boltzmanneq}
\eea
where $\lambda$ is a dimensionless constant that denotes the strength of the collisions, and \begin{equation}
    \int_{k} = \int \frac{d^3\mathbf{k}}{(2\pi)^3 E_k},
\end{equation}
is the Lorentz invariant momentum space integral \cite{degroot}. We note that equilibrium distribution function is given by 
\be
f_k^{eq}=\alpha\, e^{-E_k},
\label{define_equilibrium}
\ee
where $\alpha>0$ is the fugacity, and this function is a zero of the collision term, i.e. $\mathcal{C}[f_k^{eq}]=0$ \cite{degroot}.

In the next section we develop a covariant generating function method and use it to determine the exact set of eigenvalues and eigenfunctions of the scalar part of the spectrum of the linearized collision operator associated with \eqref{Boltzmanneq}. We then show in Section \ref{moments_method} that the method can also be used to determine the exact set of equations of motion for suitably defined scalar moments of the distribution function, which we solve to determine the corresponding solution of the Boltzmann equation.

\section{Scalar spectrum of the linearized collision operator}
\label{linearized_soln}

The linearized Boltzmann equation is framed in terms of perturbations about the equilibrium distribution function, truncated at first order in deviations. 
The distribution function is written as
\begin{equation}
    f_k = f_{k}^{eq} + \delta f_k \equiv f_{k}^{eq}(1 + \phi_k),
\end{equation}
where $\phi_k$ parameterizes the deviations from equilibrium. After substituting this expression into the Boltzmann equation \eqref{Boltzmanneq}, the result to linear order in $\phi_k$ is
\begin{equation}
    E_k \partial_{\tau} (\phi_k f_{k}^{eq}) = \frac{\lambda}{2} \int_{k'pp'} (2\pi)^5  \delta^{(4)}(k_{\mu} + k'_{\mu} - p_{\mu} - p'_{\mu}) f_k^{eq}f_{k'}^{eq} (\phi_{p'} + \phi_p - \phi_{k'} - \phi_k).
\end{equation}
Defining the linearized Boltzmann collision operator as
\begin{equation}
    \mathcal{L}[\phi] \equiv \frac{\lambda}{2} \int_{k'pp'} (2\pi)^5 \delta^{(4)}(k_{\mu} + k'_{\mu} - p_{\mu} - p'_{\mu}) f_{k'}^{eq} (\phi_{p'} + \phi_p - \phi_{k'} - \phi_k),
\end{equation}
the linearized Boltzmann equation can then be written as
\begin{equation}
    E_k \partial_{\tau} (\phi_k f_{k}^{eq}) = f_{k}^{eq} \mathcal{L}[\phi_k] = f_k^{eq} \left( \mathcal{L}_{gain}[\phi_k] - \mathcal{L}_{loss}[\phi_k] \right),
\end{equation}
where we have decomposed the linearized collision operator into a gain term and a loss term which are respectively given by
\begin{equation}
    \mathcal{L}_{gain}[\phi_k] \equiv \frac{\lambda}{2} \int_{k'} f_{k'}^{eq} \int_{pp'} (2\pi)^5 \delta^{(4)}(k_{\mu} + k'_{\mu} - p_{\mu} - p'_{\mu}) (\phi_p + \phi_{p'}),
\end{equation}
\begin{equation}
    \mathcal{L}_{loss}[\phi_k] \equiv \frac{\lambda}{2} \int_{k'} f_{k'}^{eq} (\phi_k + \phi_{k'})\int_{pp'} (2\pi)^5 \delta^{(4)}(k_{\mu} + k'_{\mu} - p_{\mu} - p'_{\mu}) .
\end{equation}

To determine the spectrum of the linearized Boltzmann operator, we propose an Ansatz for the eigenfunctions in terms of the associated Laguerre polynomial, $L_n^{(1)}(E_k)$ \cite{gradshteyn2007}. We note that 
\be
\int_{pp'} (2\pi)^5 \delta^{(4)}(k_{\mu} + k'_{\mu} - p_{\mu} - p'_{\mu})=1
\label{ppprime_integral}
\ee
as shown in \cite{Bazow:2015dha}, so $\mathcal{L}_{loss} = \frac{\lambda}{2} \int_{k'} f_{k'}^{eq} (\phi_k + \phi_k')$. Substituting $\phi_k=L_n^{(1)}$, one finds  
\begin{equation}
    \mathcal{L}_{loss} = \frac{\lambda}{2}L_n^{(1)}(E_k) \int_{k'} f_{k'}^{eq} + \frac{\lambda}{2} \int_{k'} f_{k'}^{eq} L_n^{(1)}(E_{k'}).
\end{equation}
The first integral can be evaluated explicitly to obtain $\int_k f_{k}^{eq} = \frac{\alpha}{2\pi^2}$. The second integral must be zero because of the orthogonality of the associated Laguerre polynomials. Therefore, the loss term becomes
\begin{equation}
    \mathcal{L}_{loss} \left[ L_{n}^{(1)}(E_k) \right] = \frac{\alpha\lambda}{4\pi^2} \left( 1 + \delta_{n0} \right) L_n^{(1)}(E_k).
\end{equation}
This result indicates that, if $L_n^{(1)}$ is an eigenfunction of $\mathcal{L}$, it will be an eigenfunction of both $\mathcal{L}_{loss}$ and $\mathcal{L}_{gain}$, separately.

The gain term is considerably more difficult to investigate. For this purpose, it is convenient to use the generating function for the associated Laguerre polynomials [see \eqref{generating_function_definition_Laguerre}] 
\begin{equation}
    \frac{1}{(1-v)^2} \exp \left( -\frac{xv}{1-v} \right) = \sum_{n=0}^{\infty} v^n L_n^{(1)}(x),
    \label{Laguerre_Generator}
\end{equation}
with $v\in [0,1)$. Further properties of this generating function and the associated Laguerre polynomials are discussed in Appendix A. Motivated by seminal work of Refs.\ \cite{Bobylev,KrookWu1976,KrookWu1977}, we define the following quantity
\begin{equation}
    I = \sum_{n=0}^{\infty} v^n \mathcal{L}_{gain}[L_n^{(1)}],
\end{equation}
which can also be written as
\begin{equation}
    I = \frac{\lambda}{(1-v)^2} \int_{k'} f_{k'}^{eq} (2\pi)^5 \int_{pp'} \delta^{(4)}(P_{T\,\mu} - p_{\mu} - p'_{\mu}) \exp\left( -E_p \frac{v}{1-v} \right),
    \label{define_I}
\end{equation}
where we defined the total 4-momentum $P_T^\mu = k^{\mu} + k'^{\mu}$. 

Since the integral over $pp'$ is a Lorentz scalar that depends only on the timelike 4-vectors $u^{\mu}$ and $P_T^{\mu}$, the result of these integrals can only depend on those 4-vectors via $u_\mu P_T^\mu$ and $s=P_{T\mu} P_T^\mu$, with $s$ being the traditional Mandelstam variable \cite{degroot}. This implies that this integral is invariant under the exchange of $u^{\mu}$ and the normalized total 4-momentum,  $\hat{P}^{\mu}_T \equiv P_T^{\mu}/\sqrt{s}$. This interchange leads to
\begin{equation}
I = \frac{\lambda}{(1-v)^2} \int_{k'} f_{k'}^{eq} (2\pi)^5 \int_{pp'} \delta^{(4)}(\sqrt{s}u_{\mu} - p_{\mu} - p'_{\mu}) \exp\left( -\hat{P}_T^\mu p_\mu \frac{v}{1-v} \right).
\end{equation}
The integral $I$ is a Lorentz scalar and, therefore, can be calculated in any Lorentz frame. For the sake of convenience, we perform this task in the local rest frame of the system where $u^\mu = (1,0,0,0)$. In this frame, $\delta^{(4)}(\sqrt{s}u_\mu-p_\mu -p'_\mu) = \delta(\sqrt{s}-E_p-E_{p'})\delta^{(3)}(\mathbf{p}+\mathbf{p'})$. Then, the integral simplifies to
\bea
I &=& \frac{\lambda}{(1-v)^2} \int_{k'} f_{k'}^{eq} \int \frac{dp \,d\hat{\mathbf{p}}}{2\pi} \delta(\sqrt{s} - 2p) \exp\left[ - \frac{pv}{1-v} (\hat{P}_T^0 - \hat{\mathbf{P}}_T \cdot \hat{\mathbf{p}}) \right] \\
&=& \frac{\lambda}{(1-v)^2} \int_{k'} f_{k'}^{eq} \int \frac{d\hat{\mathbf{p}}}{4\pi} \exp\left[ - \frac{v}{2(1-v)} (P_T^0 - \mathbf{P}_T \cdot \hat{\mathbf{p}}) \right]\, ,
\eea
where $\hat{\mathbf{p}}=\mathbf{p}/E_p=\mathbf{p}/p$ is a unit vector in three dimensions. We further define $\cos\theta_{kp} = x$ and $\cos\theta_{k'p} = y$, where $\theta_{kp}$ is the angle between $\mathbf{k}$ and $\mathbf{p}$, and $\theta_{k'p}$ is the angle between $\mathbf{k}'$ and $\mathbf{p}$, respectively. This allows $I$ to be written as 
\begin{eqnarray}
I &=& \frac{\lambda}{(1-v)^2} \int_{k'} f_{k'}^{eq} \int_{-1}^{1} \frac{dx}{2} \exp\left[ - \frac{v}{2(1-v)} E_k(1-x)  \right] \int_{-1}^{1} \frac{dy}{2} \exp\left[ - \frac{v}{2(1-v)} E_{k'}(1-y)  \right].
\label{resultforI}
\end{eqnarray}
This integral can now be evaluated using standard techniques to find that
\begin{equation}
    I = \frac{\alpha\lambda}{v}\frac{1}{2 \pi^2 E_k} \Big[ 1 - \exp\Big( -\frac{E_k v}{1-v} \Big) \Big].
\end{equation}
Expanding term by term in powers of $v$, the gain term is then given by
\begin{equation}
    \mathcal{L}_{gain}\Big[ L_n^{(1)}(E_k) \Big] = \frac{\alpha\lambda}{2 \pi^2 (n+1)} L_n^{(1)}(E_k).
\end{equation}
 The gain term and the loss term are then combined to obtain
\begin{equation}
    \mathcal{L} \Big[ L_n^{(1)}(E_k) \Big] = \chi_n L_n^{(1)}(E_k),
\end{equation}
where the eigenvalues are given by
\begin{equation}
    \chi_n = - \frac{\alpha\lambda}{4\pi^2} \left(\frac{n-1}{n+1}+\delta_{n0}\right).
\end{equation}
 Therefore, we see that $L_n^{(1)}(E_k)$ and $\chi_n$ are, respectively, the \emph{exact} eigenfunctions and eigenvalues of the scalar part of the spectrum of the collision operator for a massless gas of particles with quartic self-interactions.  As $n \rightarrow \infty$, the eigenvalues approach $-\frac{\alpha\lambda }{4\pi^2}$, which indicates that the lifetime of the higher-order modes approaches $\frac{4\pi^2}{\alpha\lambda }$. Finally, we note that going back to standard units where the momenta are not scaled by $T_0$, one finds  $L_n^{(1)}(E_k/T_0)$ and that the eigenvalue $\chi_n$ has dimensions of $T_0^2$.

\section{Exact equations of motion for the moments}
\label{moments_method}

The goal of this section is to rewrite the full Boltzmann equation in \eqref{Boltzmanneq} in terms of ordinary differential equations for suitably defined Lorentz scalar moments of $f_k$, which can be  solved using standard numerical routines. Having in mind the results from the previous section, it is natural to expand the distribution function in terms of an associated Laguerre basis $L_n^{(\beta)}(E_k)$. However, while  $L_n^{(1)}(E_k)$ are the eigenfunctions of the \emph{linearized} collision operator, it turns out that in order to find the exact equations of motion for the \emph{nonlinear} case, it is best to consider a basis in terms of $L_n^{(2)}(E_k)$. This can be understood as follows. 

Consider the Boltzmann equation \eqref{Boltzmanneq} and multiply it on both sides by $L_n^{(\beta)}(E_k)$ (with arbitrary $\beta$) and then integrate it over $k$, which gives
\bea
&&D\int_k E_k\, L_n^{(\beta)}(E_k)\,f_k=  \frac{\lambda}{2}\,\int_{kk'pp'}L_n^{(\beta)}(E_k)
\,(2\pi)^5\,\,\delta^{(4)}(k_\mu+k'_\mu-p_\mu -p'_\mu)(f_p f_{p'}{-}f_k f_{k'}).
\label{Boltzmanneq_new}
\eea
While the right-hand side can be simplified using the techniques discussed in the previous section for any integer value of $\beta$, a bad choice for the coefficient $\beta$ above can make the left-hand side unnecessarily complex. Indeed, if $\beta=1$, after performing the angular integrals the left-hand side becomes (apart from constant multiplicative factors)
\be
D\int_0^\infty dk\, k^2 L_n^{(\beta)}(k)f_k.
\label{lhs1}
\ee
One could then decompose the distribution function in terms of generic associated Laguerre polynomials as follows
\begin{equation}
    f_k(\tau) = e^{-k} \sum_{m=0}^{\infty} c_m^{(\gamma)}(\tau) L_m^{(\gamma)}(E_k),
    \label{distribution_expansion}
\end{equation}
which can be used back in \eqref{lhs1} to find that the left-hand side becomes
\be
\sum_{m=0}^\infty Dc_m^{(\gamma)}  \,\int_0^\infty dk\, k^2\,e^{-k} L_n^{(\beta)}(k)L_m^{(\gamma)}(k).
\label{lhs2}
\ee
To avoid having a sum of terms already on the left-hand side of the equations for the moments, it is clear that one should set $\beta=\gamma=2$. Any other choice would severely complicate our analysis. In fact, in this case the orthogonality condition for the associated Laguerre polynomials \eqref{orthogonality} can be used and the moment equations become simply
\be
Dc_n = \frac{\lambda}{n_0}\frac{1}{(n+1)(n+2)}\,\int_{kk'pp'}L_n^{(2)}(E_k)
\,(2\pi)^5\,\,\delta^{(4)}(k_\mu+k'_\mu-p_\mu -p'_\mu)(f_p f_{p'}{-}f_k f_{k'}), 
\label{lhs3}
\ee
where we defined $c_n \equiv c_n^{(2)}$ to ease the notation. In this way, the moments are finally defined as in \cite{Bazow:2016oky}
\begin{equation}
    c_n(\tau) =  \frac{2}{(n+1)(n+2)n_0}\int_k E_k L_n^{(2)}(E_k) f_k(\tau)
    \label{cn_moments}
\end{equation}
and, once \eqref{lhs3} is solved and the $c_n$ moments are found, one can always recover back the distribution function as follows:
\begin{equation}
    f_k(\tau) = f_{k}^{eq} \sum_{n=0}^{\infty} c_n(\tau) L_n^{(2)}(E_k).
    \label{distribution_expansion}
\end{equation}
We note that, due to the conservation laws, $c_0=1$ and $c_1=0$ at all times. 
We will now proceed to express the right-hand side of \eqref{lhs3} directly in terms of the $c_n$ moments. The result of the integral
\be
\mathcal{J}_n=\int_{kk'pp'}L_n^{(2)}(E_k)
\,(2\pi)^5\,\,\delta^{(4)}(k_\mu+k'_\mu-p_\mu -p'_\mu)(f_p f_{p'}{-}f_k f_{k'})
\label{defineJn}
\ee
can be found using the generating function of the associated Laguerre polynomials as follows. We define 
\be
\mathcal{J}=\mathcal{J}_{\mathrm{gain}} -\mathcal{J}_{\mathrm{loss}},
\ee
where we introduced generating function related to the gain and loss terms of the collision term,
\begin{eqnarray}
\mathcal{J}_{\mathrm{gain}} &=& \frac{1}{(1-v)^3}\int_{kk'}f_k f_{k'}
\,(2\pi)^5\,\int_{pp'}\,\delta^{(4)}(k_\mu+k'_\mu-p_\mu -p'_\mu)\exp\left(-\frac{v E_p}{1-v}\right), \\
\mathcal{J}_{\mathrm{loss}} &=& \frac{1}{(1-v)^3}\int_{kk'}f_k f_{k'}
\,\exp\left(-\frac{v E_k}{1-v}\right),
\end{eqnarray}
in such a way that 
\be
\mathcal{J} = \sum_{n=0}^\infty v^n \mathcal{J}_n.
\label{gen_funct_expand}
\ee
This provides a way to determine the integral in \eqref{defineJn}.

The loss term is simpler and, thus, is evaluated first. Using the trivial identity 
\be
\frac{1}{E_k} = \int_0^{\infty} da\, e^{-a E_k} ,
\label{1/E_trick}
\ee
the loss term can be re-written as 
\be
\begin{split}
\mathcal{J}_{\mathrm{loss}} & = \frac{1}{(1-v)^3} \int_0^{\infty} da \int_0^{\infty} db \int_k f_k E_k \exp \left[ -E_k \left( \frac{v}{1-v} + a \right) \right] \int_{k'} f_{k'} E_{k'} \exp \left[ -E_{k'} \left( \frac{v}{1-v} + b \right) \right] .
\end{split}
\ee 
These exponentials can be written in terms of associated Laguerre polynomials using (for $a \geq 0$)
\be
e^{-a E_k } = \sum_{n=0}^{\infty} \frac{a^n}{(1+a)^{n+3}} L_n^{(2)}(E_k) ,
\label{exponential_expansion}
\ee 
so the loss term becomes
\be
\mathcal{J}_{\mathrm{loss}} = \sum_{n,m=0}^{\infty} \int_k f_k E_k \int_{k'} f_{k'} E_{k'} \int_0^{\infty} da \int_0^{\infty} db \frac{\left( \frac{v}{1-v} + a \right)^n}{\left( 1 + \frac{v}{1-v} + a \right)^{n+3}} \frac{\left( \frac{v}{1-v} + b \right)^m}{\left( 1 + \frac{v}{1-v} + b \right)^{m+3}} L_n^{(2)}(E_k) L_m^{(2)}(E_{k'}) .
\ee 
We then evaluate the integrals over $a,b$ and use \eqref{cn_moments} to write the loss term as 
\be
\mathcal{J}_{\mathrm{loss}} = \frac{n_0^2}{4(1-v)^3} \sum_{n,m=0}^{\infty} c_n c_m \left[ \left( nv - n + v - 2 \right) v^{n+1} + 1 \right] .
\ee

To compute the gain term, we first evaluate the integral 
\be
\begin{split}
    \mathcal{P} & = \int_{pp'} \delta^{(4)}(k_{\mu} + k_{\mu}' - p_{\mu} - p_{\mu}') \exp \left( -\frac{vE_p}{1-v} \right) \\
    & = \exp \left[ -\frac{P_T v}{2(1-v)} \right] \int_{pp'} \exp \left[ \frac{1}{2} (E_p - E_{p'}) \frac{v}{1-v} \right] \delta^{(4)}(P_T - p - p') \\
    & = \exp \left[ -\frac{P_T v}{2(1-v)} \right] \int_{pp'} \exp \left[ \frac{1}{2} \hat{\mathbf{P}}_T \cdot (\mathbf{p} - \mathbf{p}') \frac{v}{1-v} \right] \delta^{(4)}(\sqrt{s} u - p - p') ,
    \label{define_P}
\end{split}
\ee
where we have once again used the symmetries of the integral to switch $u^{\mu} \leftrightarrow \hat{P}^\mu_T$, as was done when deriving the eigenfunctions of the linearized collision operator. We thus arrive at 
\be
\begin{split}
    \mathcal{P} & = \frac{1}{(2\pi)^5} \exp \left[ -\frac{(E_k+E_{k'}) v}{2(1-v)} \right] \int_0^{\infty} dp \int \frac{d\hat{\mathbf{p}}}{4\pi} \delta \left( \frac{\sqrt{s}}{2} - E_p \right) \exp \left( -\frac{v E_p}{1-v} \hat{\mathbf{P}}_T \cdot \hat{\mathbf{p}} \right) \\
    & = \frac{1}{(2\pi)^5} \exp \left[ -\frac{(E_k+E_{k'}) v}{2(1-v)} \right] \int \frac{d\hat{\mathbf{p}}}{4\pi} \exp \left[ -\frac{v}{2(1-v)} \mathbf{P}_T \cdot \hat{\mathbf{p}} \right] .
    \label{P_result}
\end{split}
\ee 
The gain term is then given by 
\be
\mathcal{J}_{\mathrm{gain}} = \frac{1}{(1-v)^3} \int_{kk'} f_k f_{k'} \exp \left[ -\frac{(E_k+E_{k'}) v}{2(1-v)} \right] \int \frac{d\hat{\mathbf{p}}}{4\pi} \int \frac{d\hat{\mathbf{k}}}{4\pi} \int \frac{d\hat{\mathbf{k}}'}{4\pi} \exp \left( -\frac{v}{2(1-v)} \mathbf{P}_T \cdot \hat{\mathbf{p}} \right) .
\ee 

As in the linearized case, we define $x = \cos\theta_{kp}$ and $y = \cos\theta_{k'p}$ so that the gain term becomes 
\be
\mathcal{J}_{\mathrm{gain}} = \frac{1}{4(1-v)^3} \int_{kk'} f_k f_{k'} \exp \left[ -\frac{(E_k+E_{k'}) v}{2(1-v)} \right] \int_{-1}^1 dx \int_{-1}^1 dy \exp \left[ -\frac{1}{2} \frac{v}{1-v} (E_k x + E_{k'} y) \right] .
\ee 
By defining $X = \frac{1}{2} \frac{v}{1-v} (1+x)$ and $Y = \frac{1}{2} \frac{v}{1-v} (1+y)$, and using \eqref{1/E_trick}, we obtain
\begin{eqnarray}
    \mathcal{J}_{\mathrm{gain}} = \frac{1}{v^2 (1-v)} \int_0^{v/(1-v)} dX \int_0^{v/(1-v)} dY \int_0^{\infty} da \int_0^{\infty} db \int_k f_k e^{-E_k (X+a)} \int_{k'} e^{- E_{k'} (Y + b)} .
\end{eqnarray}
Once again using \eqref{exponential_expansion}, this can be expressed as
\be
\begin{split}
    \mathcal{J}_{\mathrm{gain}} = & \frac{1}{v^2 (1-v)} \sum_{n,m = 0}^{\infty} \int_k f_k E_k L_n^{(2)}(E_k) \int_{k'} f_{k'} E_{k'} L_m^{(2)}(E_{k'}) \times \\
    & \times \int_0^{v/(1-v)} dX \int_{0}^{\infty} da \frac{(X+a)^n}{(1+X+a)^{n+3}} \int_0^{v/(1-v)} dY \int_{0}^{\infty} db \frac{(Y+b)^m}{(1+Y+b)^{m+3}} .
\end{split}
\ee 
Evaluating the integrals over $a,b,X,Y$ using standard techniques and combining this with \eqref{cn_moments}, we find that the gain term is given by 
\be
\mathcal{J}_{\mathrm{gain}} = \frac{n_0^2}{4(1-v)^3} \sum_{n,m=0}^{\infty} c_n c_m \left( v^{n+1} - 1 \right) \left( v^{m+1} - 1 \right) .
\ee 
All that remains is to combine the gain and loss terms. Once that is done, one finds 
\begin{eqnarray}
    \mathcal{J}_{\mathrm{gain}} - \mathcal{J}_{\mathrm{loss}} = \frac{n_0^2}{4(1-v)^3} \sum_{n,m=0}^{\infty} c_n c_m \left[ v^{n+m+2} + n v^{n+1} - (n+1) v^{n+2} \right] .
\end{eqnarray}
After performing several redefinitions of summation indices and using the expansion  \be 
\frac{1}{(1-v)^3} = \frac{1}{2} \sum_{l=0}^{\infty} \frac{(l+2)!}{l!} v^l,
\ee 
valid for $|v|<1$, we obtain 
\be
\mathcal{J}_{\mathrm{gain}} - \mathcal{J}_{\mathrm{loss}} = \frac{n_0^2}{8} \sum_{N=2}^{\infty} v^N \sum_{n=0}^{N-2} \binom{N-n}{2} \left[ \sum_{m=0}^n c_{n-m} c_m + (n+1) (c_{n+1} - c_n) \left( \sum_{l=0}^{\infty} c_l \right) \right] .
\label{gainminloss}
\ee 
This has the same form as \eqref{gen_funct_expand} except it is missing the first two terms, $N=0,1$, which are fixed by the conservation laws. So, we use the generating function to determine that
\be 
\frac{\lambda}{n_0} \left( \mathcal{J}_{\mathrm{gain}} - \mathcal{J}_{\mathrm{loss}} \right) = \sum_{N=0}^{\infty} (N+1)(N+2) v^N D c_N ,
\ee 
where the first two terms will not contribute because $Dc_0 = Dc_1 = 0$. Comparing this to  \eqref{gainminloss} we arrive at a set of equations for the evolution of the moments,
\begin{eqnarray}
    D c_N = \frac{dc_N}{d\tau} = \frac{\lambda n_0}{8} \frac{N!}{(N+2)!} \sum_{n=0}^{N-2} \binom{N-n}{2} \left[ \sum_{m=0}^n c_{n-m} c_m + (n+1) (c_{n+1} - c_n) \left( \sum_{l=0}^{\infty} c_l \right) \right] ,
    \label{solution}
\end{eqnarray}
for $N \geq 2$. This defines a set of equations  that  determines the exact time evolution of the moments $c_N$, with which a full solution for the distribution function of the Boltzmann equation can be reconstructed using Eq.\ \eqref{distribution_expansion}. As such, the equations above can be used to determine how an arbitrarily far from equilibrium state of the gas of massless scalar particles evolves in time in an expanding FLRW universe. To the best of our knowledge, this is the first time the exact set of equations of motion for the (scalar) moments of the full nonlinear Boltzmann equation describing massless scalar particles has been derived.     

It is instructive to compare our result in \eqref{solution} to the corresponding set of equations derived in \cite{Bazow:2015dha,Bazow:2016oky} that describes a massless gas with constant cross section in FLRW. In the latter, because of the drastic assumption about the particle interactions, the solution of the n-th moment only depended on the dynamics of the previous moments. Therefore, for the system considered in \cite{Bazow:2015dha,Bazow:2016oky} an iterative procedure could be easily employed to obtain the moments for arbitrary initial conditions. Furthermore, the equations simplified so much in that case that even an analytical solution for the moments (and, consequently, to $f_k$) could be found \cite{Bazow:2015dha}. In contrast, in the case of $\lambda \phi^4$ interactions where the cross section varies with the energy, the results of this section show that one is still able to find the exact set of equations \eqref{solution} that describes the evolution of the moments, but now we see that the derivative of the n-th moment does not depend only on the previous moments. Rather, it depends on the sum over all the moments via  $\sum_{l=0}^{\infty} c_l$. This means that analytical solutions will be even harder to find than before. Furthermore, \eqref{solution} cannot be solved using a simple iterative scheme. However, \eqref{solution} can still be solved numerically, as we show in Section \ref{numerics}.

\subsubsection{Uniqueness of equilibrium}

In this section we show that the asymptotic equilibrium state solution of \eqref{solution} is unique, as expected. The argument works as follows. First assume that all the moments have reached their asymptotic state such that $dc_N/d\tau = 0$ for all $N$. Then prove using the equations of motion \eqref{solution} that this implies that all moments $c_N$ for $N \geq 2$ are equal to zero -- which indicates that the standard equilibrium state, described by the distribution $f_k^{eq}$, has been reached. 

The condition $dc_N/d\tau = 0$ implies that 
\begin{equation}
    0 = \sum_{n=0}^{N-2} {N-n \choose 2} \left[ \sum_{m=0}^n c_{n-m} c_m + (n+1) (c_{n+1} - c_n) \big( 1 + \mathcal{M} \big) \right]
\end{equation}
for all $N$. Above, we introduced the quantity 
\begin{eqnarray}
    \mathcal{M} = \sum_{N=2}^{\infty} c_N .
\end{eqnarray}
The binomial in the summation above is zero whenever $2 > N-n$, so we start by considering the case where $N = 2$. Then, the only term that will contribute to the sum is that of $n=0$, which gives the condition
\begin{equation}
    \mathcal{M} = 0.
\end{equation}
The next case to consider is that of $N = 3$. Now, two terms will contribute to the sum, those of $n=0,1$. This will then give the condition
\begin{equation}
    c_2 = 0.
\end{equation}
Using that the values of the two first moments are already known, $c_0=1$ and $c_1=0$, this condition provides an initialization for a proof by induction. 

Proceeding with standard induction, it is assumed that $c_2, ..., c_k = 0$ for some $k \geq 2$. Then, $N = 1 + k$ is considered. In this case, we have 
\begin{equation}
    0 = \sum_{n=0}^{k - 1} {k - n + 1 \choose 2} \left[ \sum_{m=0}^n c_{n-m} c_m + (n+1) (c_{n+1} - c_n) \left( 1 + \mathcal{M} \right) \right].
\end{equation}
Since we have assumed that the only non-zero $c_n$ for $n < k$ is $c_0 = 1$, it follows that the term $c_{n-m} c_m$ is only non-zero when $n-m,m$ are some combination of $0,k+1$. We also know that the binomial is zero whenever $1 > k - n + 1$, which is equivalent to $n > k$. So, it follows that $n$ can never reach $k+1$ and the only non-zero terms will occur when $n=m=0$. But, as before this will always cancel with the case of $c_n$ in the next term. So, all that remains in the summation is 
\begin{equation}
    0 = \sum_{n=1}^{k - 1} {k - n + 1 \choose 2} (n + 1) (c_{n+1} - c_n).
\end{equation}
However, the binomial is equal to zero for $n > k$, so we can really truncate this series at
\begin{equation}
    0 = \sum_{n=1}^k {k - n + 1 \choose 2} (n + 1) (c_{n+1} - c_n).
\end{equation}
This leaves only one term, 
\begin{equation}
    0 = (k + 1) c_{k+1} \rightarrow c_{k+1} = 0.
\end{equation}
By induction, it follows that all $c_n$ are equal to zero except for $c_0 = 1$. This means that the only steady-state solution is the standard equilibrium state, as expected. This provides an alternative way to show that the equilibrium state is unique, starting from the exact moment equations.

\subsubsection{A comment on the thermalization time}\label{comment}

In the following, we will show that $\mathcal{M}$ only vanishes in equilibrium, i.e., when all moments $c_{m\geq 2}=0$. First, we assume that $\mathcal{M}$ is equal to zero for all $\tau \geq \tau_0$. Using this assumption, the equation of motion for the second moment obtained from \eqref{solution}, evaluated  at $\tau \geq \tau_0$, implies that $dc_2/d\tau = 0$ and, consequently, that $c_2$ is a constant. If we then analyze the equation of motion for the third moment, we see that $dc_3/d\tau$ only depends on $c_2$ and, therefore, this derivative is a constant. This would imply that $c_3$ is a linear function of time. Similarly, if one analyses the equation of motion for the fourth moment, it will depend only on $c_2$ and $c_3$ and, thus, $c_4$ will be a quadratic function of time. The same argument can be applied to the remaining moments and one will conclude that the moments will all display a polynomial behaviour with time, with the highest possible power of time being $\tau^{n-2}$ for $c_n$. Therefore, this type of solution diverges as $\tau \to \infty$, which is incompatible with the existence of the equilibrium state and the fact that moments of $f_k$ must be finite. The only solution that does not display this divergent behavior corresponds to the case where $c_{n\geq 2}=0$ -- the equilibrium solution. Therefore, one can see that the thermalization time of the system is well estimated by the time it takes for  $\mathcal{M}$ to vanish.

\begin{figure}[!h]
    \centering
    \includegraphics[width=1.0\linewidth]{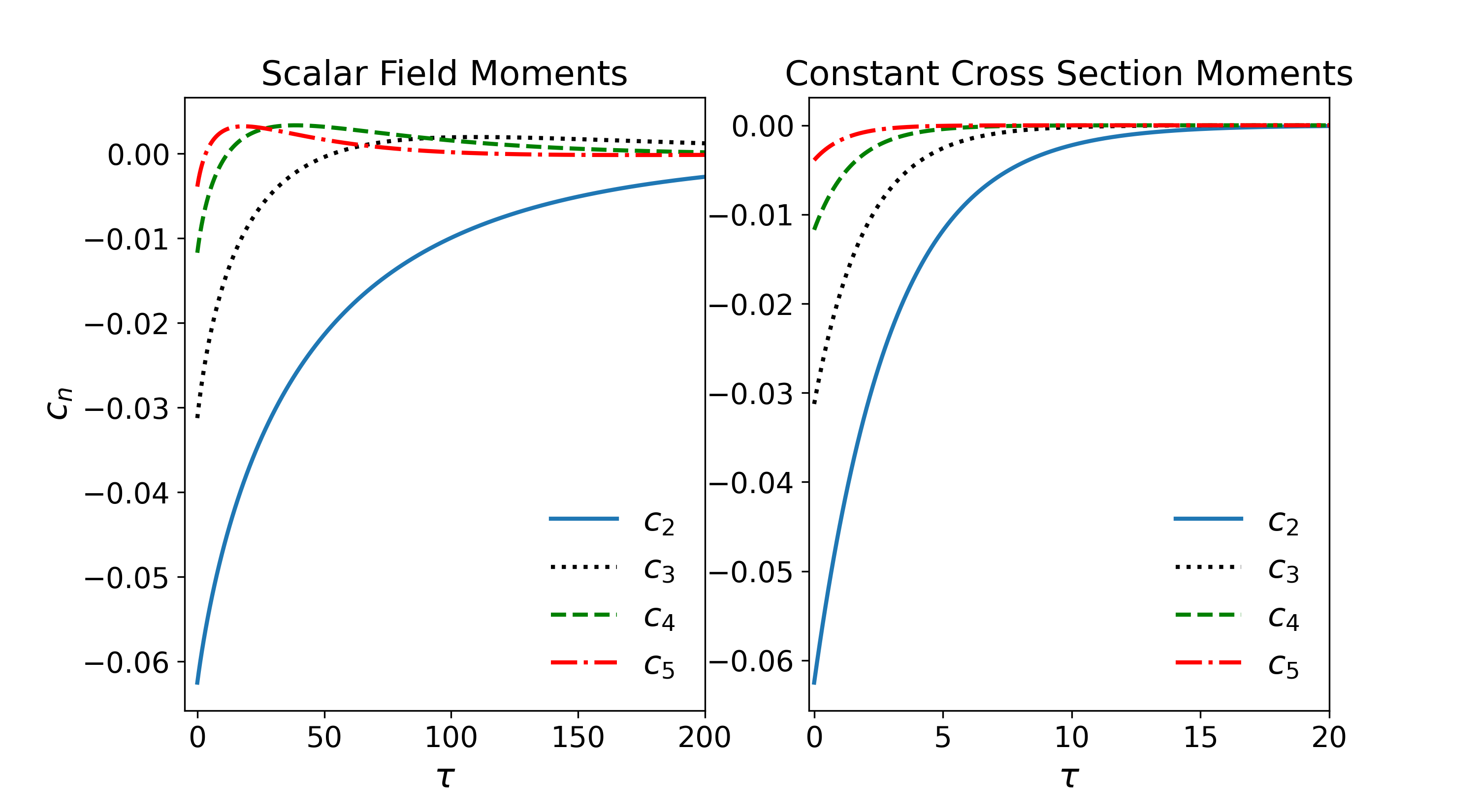}
    \caption{Comparison of the evolution of the first four non-constant moments between the scalar field case (left), and the constant cross section case (right).}
    \label{analytical_moments}
\end{figure}

\section{Numerical results and comparison}
\label{numerics}

\begin{figure}[!h]
    \centering
    \includegraphics[width=1.0\linewidth]{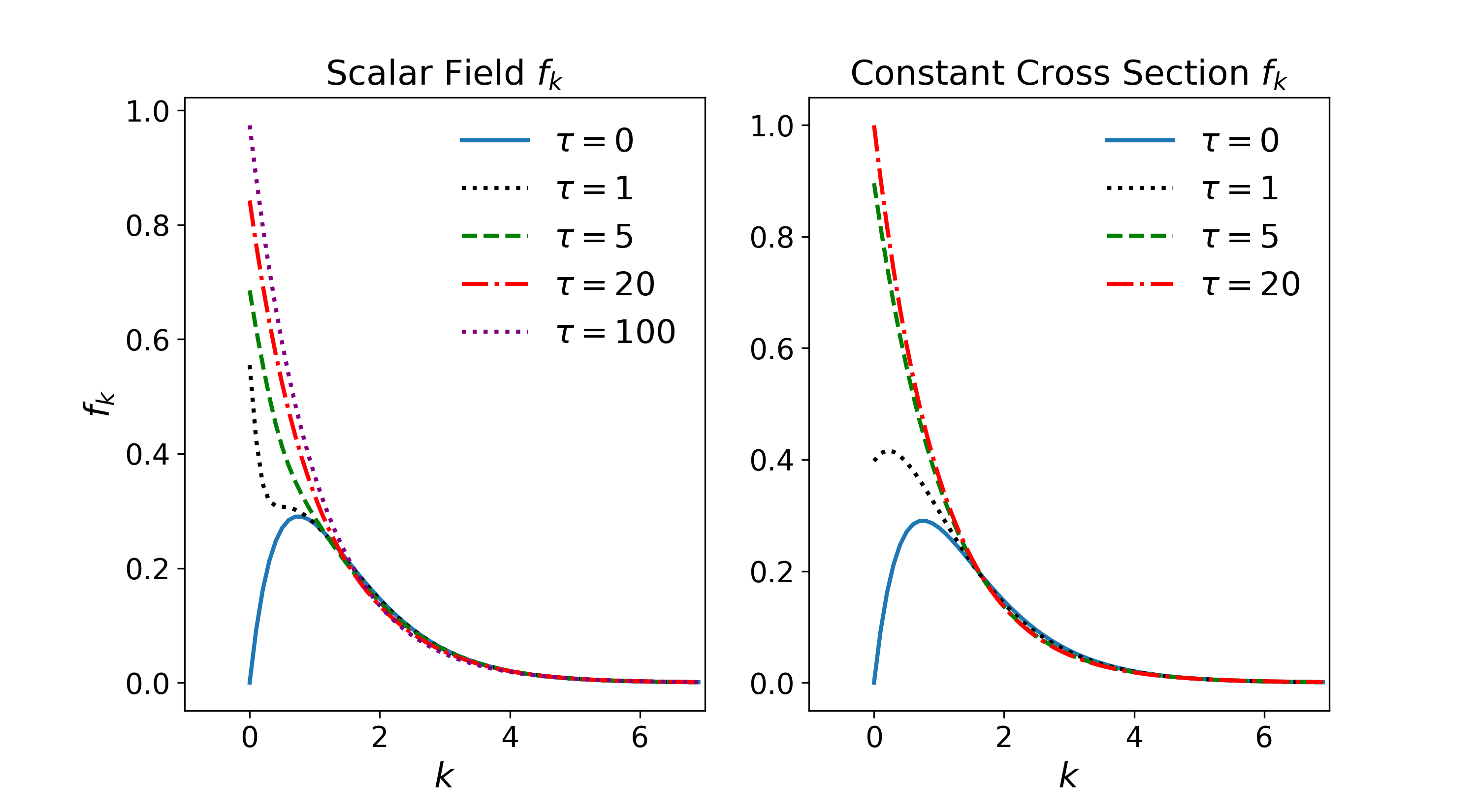}
    \caption{Comparison of the evolution of the distribution function between the scalar field case (left), and the constant cross section case (right).}
    \label{dist_funct}
\end{figure}

\begin{figure}[!h]
    \centering
    \includegraphics[width=1.0\linewidth]{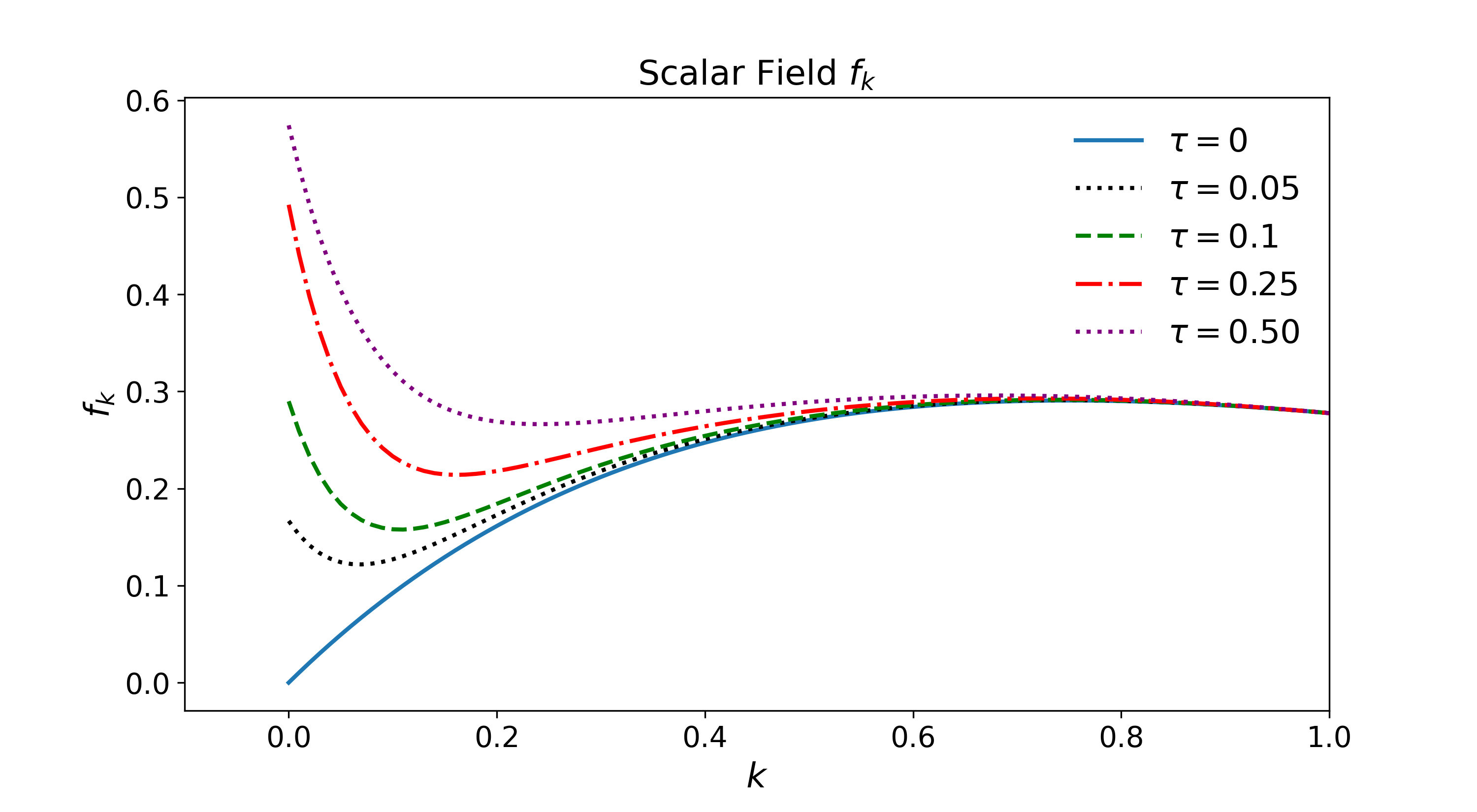}
    \caption{Early time evolution of the scalar field distribution function.}
    \label{early_time}
\end{figure}

Since an exact analytical solution has not been obtained, simple numerical procedures are required to solve the moment equations and compare their solutions to those found in the constant cross section case previously considered in \cite{Bazow:2015dha,Bazow:2016oky}. The evolution equation for the moments in the scalar field case contains an infinite summation, which makes it harder to solve in comparison to the constant cross section case. Here solutions are obtained by considering some maximum moment defined by $N_{max}$ and numerically solving the evolution equations up to that highest moment. To choose this maximum moment, it is examined at which point the evolution of the moments and the distribution function reach a steady state and do not appreciably change anymore. For each of the figures presented in this work, the maximum moment $N_{max} = 90$ is used, and the time is scaled by $\lambda n_0$. To solve the differential equation after this cutoff is applied, a fourth-order Runge-Kutta algorithm is applied. 

Our results are compared to the constant cross section case so a brief summary of the results found in the latter is included here. In \cite{Bazow:2015dha, Bazow:2016oky} an exact differential equation for the moments was derived for the case of massless particles interacting with a constant cross section. However, the differential equation for the n-th moment only depends on moments with order less than or equal to $n$. This allows for a simple iterative approach to be employed to obtain an exact analytical solution for each of the moments and, therefore, for the full distribution function. In particular, for the initial condition
\begin{equation}
    c_n(0) = \frac{1-n}{4^n},
    \label{initial_conditions}
\end{equation}
it is found that the corresponding (Laguerre-based) moments for hard spheres evolve as \cite{Bazow:2016oky}
\begin{equation}
    c_n(\tau) = \frac{1-n}{4^n} e^{-n\tau / 6}.
    \label{analytical_solution}
\end{equation}
This solution is then used as the basis for numerical simulations of the scalar field solution. In fact, we use $\eqref{initial_conditions}$ as the initial conditions for the moments satisfying the equations of motion \eqref{solution}, derived for the scalar field case. The evolution of the moments and the corresponding distribution function for the scalar field system are then compared to the analytical solution in \eqref{analytical_solution}. 

\begin{figure}[!h]
    \centering
    \includegraphics[width=1.0\linewidth]{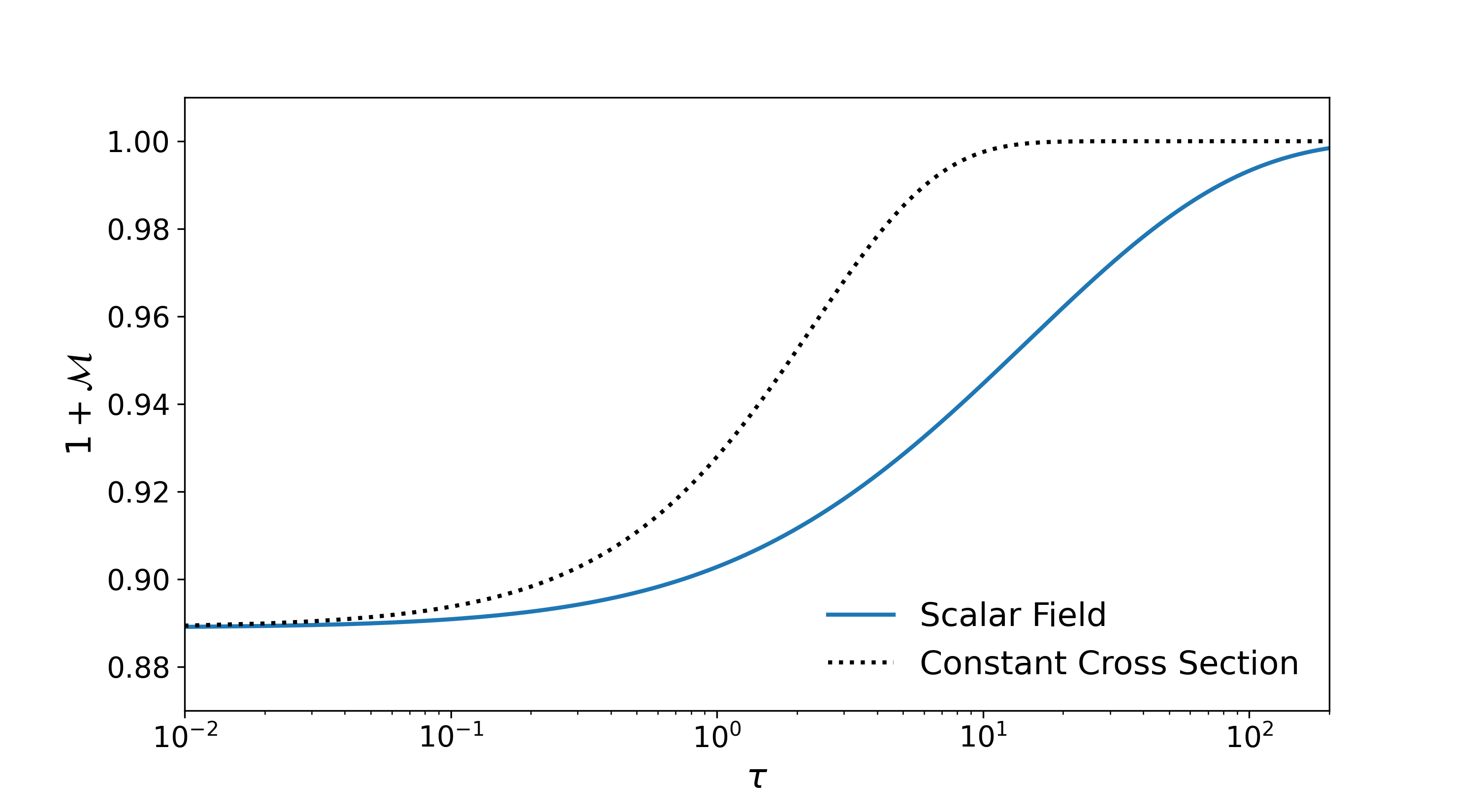}
    \caption{Comparison of the evolution of the sum of the moments $\mathcal{M}$ for the constant cross section and scalar field cases.}
    \label{debye_mass}
\end{figure}

\begin{figure}[!h]
    \centering
    \includegraphics[width=1.0\linewidth]{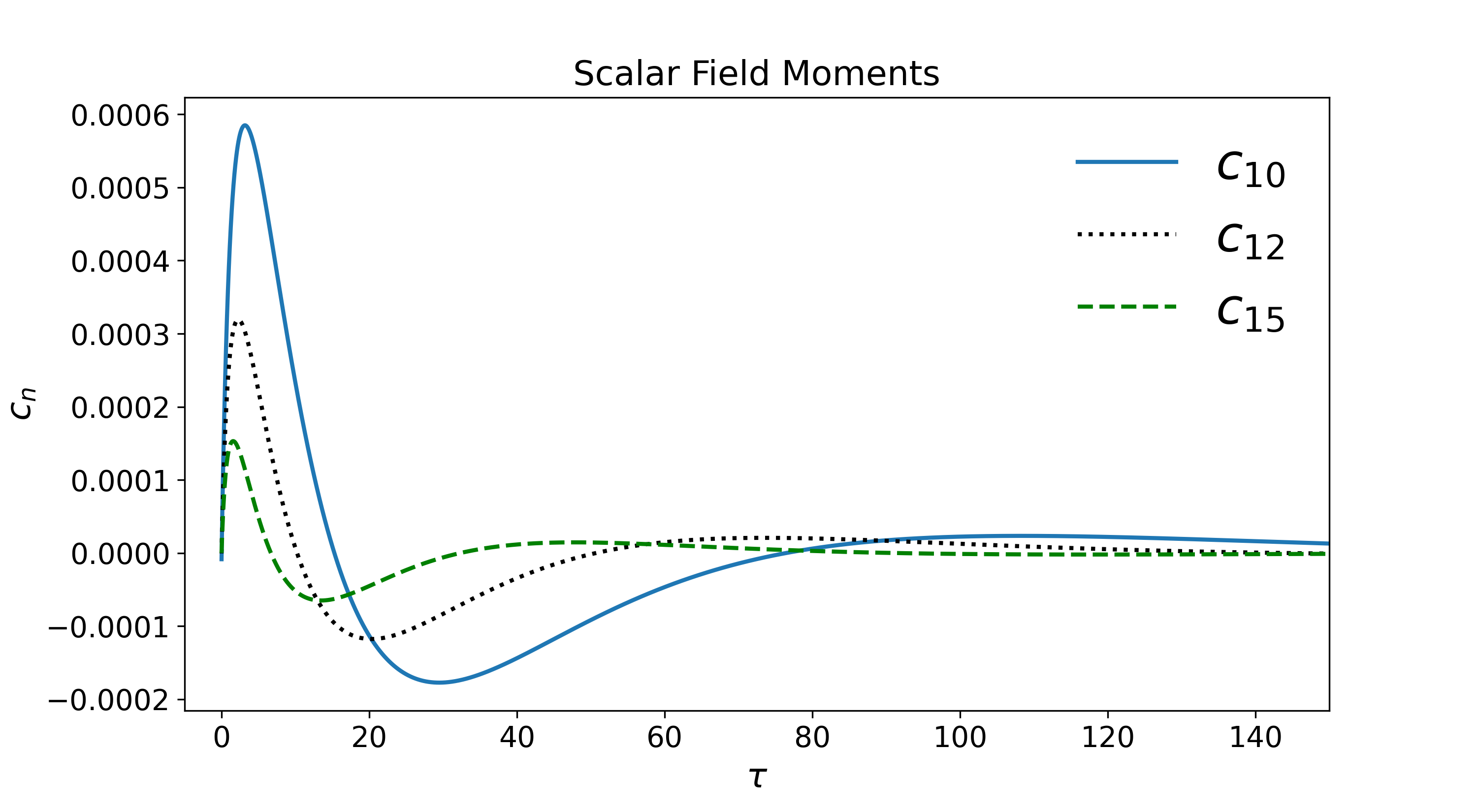}
    \caption{Evolution of some higher order moments for the scalar field case.}
    \label{high_moments}
\end{figure}

Figure \ref{analytical_moments} shows the evolution of the Laguerre moments as a function of time for the scalar field and constant cross section cases. In the constant cross section case, each moment directly approaches the equilibrium value at an exponential rate according to the analytical solution. However, in the scalar field case there is some oscillation in many of the moments. In Fig.\ \ref{dist_funct}, the effects of these differences on the distribution function can be seen. In the constant cross section case, the system equilibrates considerably faster, and also more uniformly in time. On the other hand, in the scalar field case, the system initially equilibrates faster for low values of $k$, before smoothing out as the system approaches equilibrium over time. Figure \ref{early_time} shows how the scalar field distribution function evolves for early times. It can be seen clearly that the initial equilibration process is much more rapid near zero momentum. It would be interesting to see how this is modified when quantum statistics effects are taken into account. 

In Fig.\ \ref{debye_mass} the evolution of $1 + \mathcal{M} = \sum_{n=0}^{\infty} c_n$ is examined for both interactions. We remind the reader that it was argued earlier in Section \ref{comment} that this quantity provides an estimate of the thermalization time (and rate). In the constant cross section case, it appears that the rate at which the system approaches equilibrium increases rapidly after early times, until the system almost reaches equilibrium. On the other hand, the rate at which the scalar field system approaches equilibrium initially is much slower, steadily increasing over time before slowing down again as equilibrium is approached. This discrepancy may be explained by the oscillatory behavior of moments seen in Fig.\ \ref{high_moments}. Initially some of the moments actually rapidly diverge from the equilibrium value, before more slowly oscillating back towards the equilibrium value. In other words, the scalar field moments do not monotonically approach equilibrium, unlike the constant cross section where all of the moments exponentially decay to equilibrium. 

\section{Conclusions}
\label{conclusion}

In this paper we investigated the dynamics of a gas of massless scalar particles with quartic (tree level) self-interactions in Friedmann-Lemaitre-Robertson-Walker spacetime, described by the Boltzmann equation. We demonstrated that the nontrivial far-from-equilibrium dynamics of this system can be determined by solving an infinite set of ordinary differential equations for suitably defined moments of the distribution function. Unlike the case of a gas with constant cross section considered in Refs.\ \cite{Bazow:2015dha,Bazow:2016oky}, the fact that in $\lambda\phi^4$ the cross section $\sigma(s)\sim 1/s$ makes deriving the exact set of equations of motion for the moments a much more complex task. 

We have overcome this challenge by using a new covariant generating function for the scalar moments of the Boltzmann distribution function (see \ref{linearized_soln} and \ref{moments_method}). This method, which is a relativistic generalization of the generating function techniques used in \cite{Bobylev,KrookWu1976,KrookWu1977}, was used here to analytically determine for the first time the eigenfunctions and eigenvalues of the scalar part of the spectrum of the linearized collision operator. The spectrum of the collision operator has never been determined in the relativistic regime and even results in a given subspace (such as the scalar sector) are extremely rare -- the only other known result can be found in Ref.~ \cite{Bazow:2015dha,Bazow:2016oky}, also in the scalar sector. Furthermore, this covariant generating function was also employed to find, for the first time, the exact nonlinear set of equations of motion for the scalar moments in the full nonlinear regime. We showed that the dependence of the cross section with the center of mass energy implies that moments of arbitrarily high order directly couple to low order moments. This should be compared to the constant cross case studied in \cite{Bazow:2015dha,Bazow:2016oky} where the $n-$order moment only coupled to moments of order $m<n$. 

Numerical solutions for the scalar field case were presented and compared to those found for a gas of hard spheres, for the same set of far-from-equilibrium initial conditions. Overall, we found that the dependence of the cross section with the energy introduces more structure in the time evolution of the system, with the moments in the scalar field case displaying oscillations in their approach to equilibrium, while for the constant cross section example the same moments just quickly exponentially decay towards their equilibrium values. This different behavior has consequences to the distribution function as well, which is reconstructed using the moments. In the gas with constant cross section, the system equilibrates much more quickly and also more uniformly in time. On the other hand, for the scalar field case, the system initially equilibrates much faster for low values of $k$, before smoothing out as it approaches equilibrium over time. We remark that we only considered classical (Boltzmann) statistics in this work. It would be very interesting (and challenging) to generalize our approach to the case where the bosonic nature of the particles is taken into account, as done in Ref.\ \cite{Almaalol:2018ynx}. In that context, one could investigate if our approach would be useful in the investigation of the far-from-equilibrium dynamics of a system that can Bose condense \cite{Blaizot:2011xf,Berges:2014bba,Berges:2015ixa,Berges:2020fwq}. 

Concerning the results of this paper, a clear next step would be to see if our generating function method could be used to derive not only the scalar part of the spectrum, but rather the full set of eigenvalues and eigenfunctions of the linearized collision operator considered here. The breaking of isotropy makes this task much more complex, given that now the generating function would have to produce all the scalar, vector, and tensor sectors of the spectrum. 

One may also check if the generating function method introduced here is useful when investigating the dynamics of systems of more relevance to heavy-ion collisions, such as QCD effective kinetic theory \cite{Arnold:2002zm,Kurkela:2015qoa,Almaalol:2020rnu,Du:2020zqg,Du:2020dvp,Du:2022bel} and models where the particles in the gas have a temperature-dependent mass \cite{Jeon:1995zm, Sasaki:2008fg,Chakraborty:2010fr,Bluhm:2010qf,Romatschke:2011qp,Alqahtani:2015qja,Rocha:2022fqz}. Furthermore, a better understanding of how the vector and tensor parts of the spectrum behave in our system would be relevant when studying the emergence of hydrodynamic attractors \cite{Heller:2015dha,Florkowski:2017olj,Romatschke:2017ejr} and the characterization of the far-from-equilibrium properties of kinetic theory systems \cite{Florkowski:2013lza, Florkowski:2014sfa,Denicol:2014xca,Denicol:2014tha, Denicol:2016bjh,Heller:2016rtz,Strickland:2017kux,Romatschke:2017vte,Denicol:2017lxn,Behtash:2017wqg,Blaizot:2017ucy,Denicol:2018pak,Almaalol:2018ynx,Mazeliauskas:2018yef,Behtash:2019txb,Brewer:2019oha,Blaizot:2021cdv,Heller:2021oxl,Chattopadhyay:2021ive,Jaiswal:2021uvv,Rocha:2022ind}. We hope to report on our progress in some of the topics mentioned above in the near future.

\section*{Acknowledgments} NM and JN are  supported by the U.S. Department of Energy, Office of Science, Office for Nuclear Physics under Award No. DE-SC0021301. GSD thanks CNPq and
Funda\c c\~ao Carlos Chagas Filho de Amparo \`a Pesquisa
do Estado do Rio de Janeiro (FAPERJ), process No. E-
26/202.747/2018, for support. JN and GSD thank the S\~ao Paulo Research
Foundation (FAPESP) under project 2017/05685-2 for support.

\appendix 
\section{Some properties of the associated Laguerre polynomials}
\label{Laguerre_properties}

The associated Laguerre polynomials are the solutions of the ordinary differential equation  \cite{gradshteyn2007}
\begin{equation}
    x \frac{d^2g(x)}{dx^2} + (\beta + 1 - x) \frac{dg(x)}{dx} + m g(x) = 0,
\end{equation}
where $\beta\neq 0$ and $m$ is a non-negative integer, and they are denoted as $L_m^{(\beta)}(x)$. These polynomials obey the orthogonality condition
\begin{equation}
    \int_0^{\infty} dx\, x^2\, e^{-x} L_n^{(\beta)}(x) L_m^{(\beta)}(x) = \frac{(n+\beta)!}{n!} \delta_{mn}.
    \label{orthogonality}
\end{equation}
The generating function for the Laguerre polynomials is given by 
\begin{equation}
    \frac{1}{(1-v)^{\beta+1}} \exp \left( -\frac{xv}{1-v} \right) = \sum_{n=0}^{\infty} v^n L_n^{(\beta)}(x).
    \label{generating_function_definition_Laguerre}
\end{equation}

\section{Generating function for constant cross section case}
\label{const_cross_soln}

While the relativistic Boltzmann equation has already been solved for constant cross section interactions in \cite{Bazow:2015dha, Bazow:2016oky}, here the moment equations are found using the new techniques presented in the main text. For this case, the only change is that the collision operator is now given by 
\begin{equation}
    \mathcal{C}[f] = \frac{\lambda}{2} \int_{k'pp'} (2\pi)^5 s\, \delta^{(4)}(k_{\mu} + k_{\mu}' - p_{\mu} - p_{\mu}') (f_p f_{p'} - f_k f_{k'}).
    \label{collision_constant_cross_section}
\end{equation}
This differs from the scalar field case only by the presence of the Mandelstam variable $s$ in the integrand of the right-hand side of \eqref{collision_constant_cross_section} to account for the fact that the cross section is constant (here, $\lambda=\sigma$ is the total cross section). We can once again use the generating function to decompose
\begin{equation}
    \mathcal{J}_{\mathrm{gain}} = \frac{\lambda}{2} \int_{kk'} f_k f_{k'} s \int_{pp'} (2\pi)^5 \delta^{(4)}(k_{\mu} + k'_{\mu} - p_{\mu} - p'_{\mu}) \frac{1}{(1-v)^{3}} \exp \left( -\frac{E_p v}{1-v} \right) ,
    \label{N_gain_const}
\end{equation}
and
\begin{equation}
    \mathcal{J}_{\mathrm{loss}} = \frac{\lambda}{2} \int_{kk'} f_k f_{k'} \frac{s}{(1-v)^{3}} \exp \left( -\frac{E_k v}{1-v} \right) .
    \label{N_loss_const}
\end{equation}
This contains the integral 
\begin{equation}
     \int_{pp'} (2\pi)^5 \delta^{(4)}(k_{\mu} + k'_{\mu} - p_{\mu} - p'_{\mu}) \frac{1}{(1-v)^{3}} \exp \left( -\frac{E_p v}{1-v} \right) ,
\end{equation}
which is exactly the same as  \eqref{define_P}, apart from an overall factor of $1/(1-v)^3$. It can thus be solved using the exact same trick of switching $u^{\mu} \leftrightarrow \hat{P}_T^{\mu}$, as was explained in the main text.

Using this result, the gain term can now be expressed as
\begin{equation}
    \mathcal{J}_{\mathrm{gain}} = \frac{\lambda}{4} \int_{-1}^{1} dx \int_{-1}^{1} dy \frac{1 - xy}{(1-v)^{3}} \int_k f_k E_k \exp \Big[ -\frac{E_k}{2} \frac{v}{1-v} (1+x) \Big] \int_{k'} f_{k'} E_{k'} \exp \left[ -\frac{E_{k'}}{2} \frac{v}{1-v} (1+y) \right]
\end{equation}
Above, $x$ and $y$ have been defined as $\cos \theta_{kp} = x$ and $\cos\theta_{k'p} = y$, just as in the case of a scalar field. From here, the integral can be simplified using the change of variables $X = \frac{1}{2}\frac{v}{1-v}(1+x)$ and $Y = \frac{1}{2}\frac{v}{1-v}(1+y)$ and by substituting an expansion of the exponential in terms of the associated Laguerre polynomials from \eqref{exponential_expansion}. The integrals can then be evaluated either using standard techniques to express the gain term as a sum over the moments
\begin{equation}
    \mathcal{J}_{\mathrm{gain}} = \frac{\lambda n_0^2}{2} \sum_{n=0}^{\infty} v^n (n+2) \sum_{m=0}^n c_{n-m}c_m .
\end{equation}

Next, we consider the loss term, which is given by 
\begin{equation}
    \mathcal{J}_{\mathrm{loss}} = \lambda \int_{kk'} f_k f_{k'} E_k E_{k'} \frac{1}{4(1-v)^{3}} \exp \left( -\frac{E_kv}{1-v} \right) \int_{-1}^1 dx \int_{-1}^1 dy \,(1-xy) ,
\end{equation}
where the same definition for $x$ and $y$ are used as in the gain term. These integrals can be evaluated explicitly, except for the integral over $k$. This gives
\begin{equation}
    \mathcal{J}_{\mathrm{loss}} = \frac{\lambda n_0}{(1-v)^{3}} \int_k E_k f_k \exp \left( -\frac{E_k v}{1-v} \right).
\end{equation}
This integral is evaluated by expanding the exponential as in \eqref{exponential_expansion}, which gives
\begin{equation}
    \mathcal{J}_{\mathrm{loss}} = \frac{\lambda n_0^2}{2} \sum_{n=0}^{\infty} v^n (n+1) (n+2) c_n .
\end{equation}

The gain and loss terms can then be combined and compared term-by-term to obtain a differential equation for the $c_n$ moments (defined in \eqref{cn_moments})
\begin{equation}
    \frac{dc_n}{d\tau} + c_n = \frac{1}{n+1} \sum_{m=0}^n c_{n-m} c_m
\end{equation}
where the time $\tau$ has been scaled by the mean free path $l_0 = 1/n_0 \lambda$. Note that this equation implies that the derivative of the n-th moment depends only on the moments $c_{i}$ for $i < n$, so the system of equations can be easily solved  iteratively.

As was shown in \cite{Bazow:2015dha, Bazow:2016oky}, for the initial condition
\begin{equation}
    c_n(0) = \frac{1-n}{4^n}
\end{equation}
there exists an analytical solution for the moments given by
\begin{equation}
    c_n(\tau) = c_n(0) e^{-n\tau / 6}.
\end{equation}
In this case each of the moments (with $n>1$) directly approaches the equilibrium value of zero rather than exhibiting some oscillatory behavior,  as found in the scalar field case. This solution is the simplest avenue to compare the constant cross section results to the corresponding results for the scalar field case using the same initial conditions \eqref{initial_conditions}, as done in the main text.


\bibliography{References}

\end{document}